# A New Approach for a Better Load Balancing and a Better Distribution of Resources in Cloud Computing

[1] Abdellah IDRISSI and [2] Faouzia ZEGRARI

Computer Sciences Laboratory (LRI), Computer Sciences Department,
Faculty of Sciences, University Mohammed V – Rabat

*Abstract*—Cloud computing is a new paradigm where data and services of Information Technology are provided via the Internet by using remote servers. It represents a new way of delivering computing resources allowing access to the network on demand. Cloud computing consists of several services, each of which can hold several tasks. As the problem of scheduling tasks is an NP-complete problem, the task management can be an important element in the technology of cloud computing. To optimize the performance of virtual machines hosted in cloud computing, several algorithms of scheduling tasks have been proposed. In this paper, we present an approach allowing to solve the problem optimally and to take into account the QoS constraints based on the different user requests. This technique, based on the Branch and Bound algorithm, allows to assign tasks to different virtual machines while ensuring load balance and a better distribution of resources. The experimental results show that our approach gives very promising results for an effective tasks planning.

## I. INTRODUCTION

Cloud computing appears as a new computer model of the company whether private, public or hybrid. It provides a paperless technical means in term of networks, servers, and storage. The cloud computing is developed primarily through distributed computing, parallel computing and grid computing. Distributed computing allows the cloud computing, to decompose a global operation into several tasks, and then send them to processing systems. The needs of the internet users are often various and depend on the tasks. However, resource planning becomes more complex in an environment composed of heterogeneous resources, and it depends on the requirements of users. Cloud computing should then integrate the resources of heterogeneous networks to minimize the completion time of all tasks and maximize resource utilization [1]. As the Quality of service (QoS) represents standards of satisfaction of using Cloud services, it is a question to coordinate the different resources and to optimize their planning.

Cloud computing is an offer for all the ICT requirements. Defined by the NIST (National Institute of Standards and Technology) [3] as a service on demand, the cloud provides access to a shared pool of configurable computing resources. On the economic front, this new model has an important budgetary and financial impact. It is a strategy to reduce costs and maximize the return on investment. Therefore, provide more flexibility and agility will be more efficient for economy. The rest of the paper is organized as follows: the next section describes the concept of cloud computing, Section 3 presents the QoS models and its various aspects. Then, we explain principles of the Genetic Algorithm in section 4, and those of the Branch and Bound followed by our propositions in section 4. In section 5, we present experimentations and results and finally, we conclude in section 6.

## II. CONCEPT OF CLOUD COMPUTING

### A. Concept of Cloud Computing

Cloud computing is not entirely a new approach. Computing has emerged in the 60s and was marked by "computing on demand", proposed by John McCarthy which he predicted in a speech that computing will one day be a public utility [4, 5]. The 80's were stacked by concepts of virtualization. Moreover, a few years ago, the ASP model (Application Service Provider) was used to propose an application as a service [5]. Through techniques of virtualization and the grid architecture [6] which allows the rise to power of a service, the idea of john is now concretized by the emergence of cloud computing. In that way, Cloud Computing was allowed to pass from the fixed price approach to the pay-as-you-go mode, a choice payment model in the disposition of the informatics demands [7, 8].

To address the problem of performance degradation of shared resources, the cloud systems use dynamic scheduling of virtual machines with the technique of dynamic migration. However, virtualization makes possible the rapid replacement of a server in a cloud without charge or major damage. It dynamically responds when allocating resources is needed, and when adapting applications with computing resources, storage and network. This option distributes workload according to the requirements requested. The Cloud belongs to the virtualization technologies and therefore provides infrastructure and platforms on demand.

This cloud model is defined by five fundamental principles, three service models and four deployment models.

The fundamental principles are the basis of all cloud computing architecture: shared resources, elasticity, self-service, payment for the use and accelerating the speed [5].

### B. Technical models [7, 10]

The structure of the cloud consists of three layers: Infrastructure, Platform and Software.

- IaaS (Infrastructure as a Service) is the lowest layer of cloud computing. It provides all the hardware

equipment that the company can rent in remote datacenters, for the need of its applications to run the IT.

- PaaS (Platform as a Service) is the service, which provides a development environment online. PaaS is the platform of execution, deployment and development of applications.

- SaaS (Software as a Service) is the final layer of cloud. It provides applications provided on request. It is a service ready to be consumed right away, accessible via Internet.

*C. Deployment models*

The main deployment models as presented in [4, 10] are:

- Public cloud or external cloud: it is a structure managed by a provider. Services are provided to various organizations via the internet ;

- Private cloud or internal cloud: the infrastructure of this type of cloud is rented by a company and is only operated by operational units via its intranet. Its infrastructure is not mutualized ;

- Hybrid Cloud: it's a mixed structure. It combines the internal resources of private cloud with external resources of public cloud ;

- Community Cloud is implemented by specific community organizations sharing common interests.

*D. Advantages and disadvantages*

The benefits of cloud computing are many through decentralization of storage space and the pooling of IT resources. We cite some advantages as: Adaptability of resources as needed [9, 10, 11, 18, 19, 20], Applications automatically benefit from security improvements and performance, Ensure high availability of services and data, and reduce risks [12]. The major disadvantage of cloud computing is related to its security. It must maintain security of information stored in the clouds, in terms of integrity, risk of intrusion, control of the documents on their storage and geographical location

### III. MODELS QOS (QUALITY OF SERVICE)

The cloud is the lever of development to meet the needs of clients. It guarantees perfect quality of service through load balancing across several servers or data centers, and again through the implementation of procedures to restore applications and data backup in case of disaster. In effect, with the multiplication of services in the cloud, several questions can be asked about the quality of service (QoS) rendered [18, 19, 20]. The QoS refers to the ability to respond with quality to user needs and to provide a service according to the requirements in terms of response time, bandwidth, availability, etc. So, the differentiator between cloud computing offerings will be the quality of service provided. The grid concept allows the growing computing power, which need monitoring the quality of provided computing resources [13]. Scheduling of resources takes account of QoS constraints on both aspects: at the user level and the system level [14].

The scheduling problem can be modeled by these three criteria, constituting a multi-objective function defined by the weighted summation of the execution time, cost and load, as defined in [8].

$$M(x) = (\omega_1 \times \text{Time}) + (\omega_2 \times \text{Cost}) + (\omega_3 \times \text{Load}) \quad (1)$$

With : $\omega_1 + \omega_2 + \omega_3 = 1$
$\omega_i$ : weight coefficient of each indicator

The Time Indicator refers to the execution time, it comprising the processor capacity (r_cpu) and bandwidth (r_comm). We will use the same formulas expressed by the authors in [8].

$$\text{Time} = (Time_{exec}) + (Time_{comm}) \quad (2)$$

$$Time_{exec} = \text{rq}_{instruction_{count}}/\text{r}_{cpu} \quad (3)$$

$$Time_{comm} = \text{rq}_{size}/\text{r}_{comm} \quad (4)$$

The rq_instruction_count is the length of the task, and the rq_size is the size of the data file.

In calculating costs, we refer to the four aspects of cost including CPU, memory, disk and bandwidth. We can define a billing formula as follows: the CPU cost versus time.

The cost is expressed by the following equation:

$$\text{Cost} = Cost_{CPU} + Cost_{Ram} + Cost_{BW} + Cost_{Stor} \quad (5)$$

The load indicator includes three parameters, which are respectively the CPU usage (Load_cpu), the memory usage (Load_mem) and the use of the bandwidth rates (Load_br) [8].

We seek to minimize the maximum load, which corresponds to the load balancing problem.

The weighted function, which allows balancing between the utilization of CPU, memory and bandwidth is given hereafter.

$$\text{Load} = 1 - \prod_{k=1}^{3}(1 - \text{Load}_k)^{\omega_{LK}} \quad (6)$$

The variables Load_cpu, load_mem and Load_br are determined as defined in [8, 15].

Through the technology of virtualization, each computing node is defined by a set of attributes constituting the Resource Information (RI) from which the task will choose for scheduling, comprising the CPU calculation ability, the memory size, the price and the load capacity of the resource. It is expressed in [8] as follows:

RI= $\{r_{cpu}, r_{mem}, r_{stor}, r_{comm}, r_{cpu_{cost}}, r_{mem_{cost}}, r_{stor_{cost}}, \} \quad (7)$

The information of the task is composed of attributes allowing demand for resources to accomplish a task [8].

$$rq = \{rq_{cpu}, rq_{mem}, rq_{stor}, rq_{comm}, rq_{instruction_{count}}, rq_{size}\} \quad (8)$$

- r_cpu: calculation ability whose unit is MIPS;
- r_mem: indicates memory size provided by the node whose unit is MB;
- r_stor: means storage space of data provided by the node. Its unit is GB;
- r_comm: refers to capacity of data transfer that node can provide. Its unit is MB/S;
- r_cpu_cost: indicates the price of the processor ;
- r_mem_cost: indicates the price of the memory. It consider 1024 MB as calculation reference;
- r_stor_cost: indicates the price of data storage. It consider 100 GB as calculation reference;
- r_comm_cost: indicates the price of bandwidth. Its unit is 1MB/S.

Based on the multiple QoS constraints environment, task scheduling of cloud computing is to allocate tasks on the appropriate resources. We focus our research fields on the task scheduling, which is one of combinatorial optimization problems. The goal is to order the execution of operations on different virtual machines VMs in the Cloud Computing environment, so as, to minimize the execution time and cost while ensuring load balancing, as described in [8, 14]. In our case, we propose to use a Branch-and-Bound algorithm and we will compare the results with genetic algorithm. We recall hereafter the description of Genetic Algorithm and Branch-and-Bound.

### IV. GENETIC ALGORITHM IN CLOUD COMPUTING

#### A. Problem

In cloud computing, users submit different kind of tasks whose requirements can be defined according to the corresponding weights to the execution time, cost and load. Given the diversity of virtual machines hosted in the cloud makes it difficult to allocate these tasks to the appropriate machines. So, to deal with this problem, we choose some computing nodes that meet the requirements in order to form the initial population of our algorithm. Then, we run the simulation that we have programmed in Java in order to obtain a satisfactory allocation of system resources.

#### B. Analysis of the problem

The principle of genetic algorithm is to evolve an initial population of individuals, by successive generation, based on the mechanism of genetic operators: crossover, mutation and selection.

We start by creating a list of virtual machines: vmList and a list of tasks: taskList. We propose that vmList is the initial solution of the problem, (SolutionList=vmList), and will contain, for each iteration, the most suitable generation.

This allows us to choose among the individuals, those that can reproduce and can undergo at the crossing. Among the various methods of selection, we opted for the random selection. The probability of selection of each individual is 1/PopSize where PopSize represents the size of the population.

The crossing allows generating one or two children by an exchange of information between two parents. As for mutation, it aims to modify a random part of the population, causing a perturbation of the gene of the chromosome, with a low rate in order to avoid a random dispersion of the population.

#### C. Genetic algorithm

```
Begin
Function algoGenetic(MaxGeneration : int, vmList :
List<Vm>, taskList : List<Task>,  allocList :
List<Allocation>)

nbGeneration : int : declaration of a generation counter
bestAlloc ← null : intialize best chromosome
constPopulationInitiale(vmList, taskList, allocList) :
creating the initial population
tabFits : Fitness[] : initialize table tabFits to store fitness of
chromosomes

While(nbGeneration < MaxGeneration)

allMi : List<Double> : initialize list to store objective
function of each chromosome

for I ← 0 to allocListSize then         ⎫
    tabFits ← fitness(allocList(i)) ;   ⎬ calculate fitness
    allMi ← tabFits[i].getMi());        ⎪ of each
end for                                  ⎭ allocation
SortMi(allMi) : sort list allMi          ⎫ select a couple
ch1 ← selection(allMi,allocList)         ⎬ of parents: ch1
ch2 ← selection(allMi,allocList)         ⎭ and ch2

parent1 ← RandListCh1                    ⎫ distribution of
parent2 ← RandListCh2                    ⎬ each parents
                                          ⎭

childrenAlloc ← Crossing(parent1,parent2) : apply
            crossover on two parents to create the children
childFit=new Fitness(childrenAlloc) : calculate fitness of
                                                children

if(childrenAlloc.getMi () <= listAveragesSort.getLast())
    replace bad parent by child in the list allocList
EndIf

childrenMutate ←  mutation(childrenAlloc) : apply
                   mutation on created child according
                   mutation probability
SortMi(allMi) : sort list allMi
Take the index for first element of the list allMi

compareAllocation(firstChildIndex , bestAlloc) : Compare
                    the best allocation, with the allocation
```

```
        already found
Replace the older generation by new generation
MaxGeneration++ : increment the number of generation

EndWhile
EndFunction
End
```

### V. BRANCH AND BOUND IN CLOUD COMPUTING

In general, the exact methods are based on finding minimum cost solutions to solve NP-hard problems [16]. These methods are implicit enumeration techniques based on the branch and bound method. They allow exploring all branches intelligently by pruning subassemblies, which do not lead to good solutions [17].

Two bounds define this technique: upper and lower. At each vertex is associated a reduction function of the cost computed. The optimal solution with respect to solutions already found is the solution of the initial problem.

#### A. Presentation

Combined with QoS constraints and the concept of optimization, we use the techniques of Branch and Bound for the scheduling tasks problem in the cloud computing environment in order to obtain a distribution scheme satisfying resources. This approach aims to allocate the tasks on the virtual machines which are more appropriate to the requirements expressed by users. It provides better results in terms of performances and costs.

#### B. Analysis of the problem

Generally, formalizing a combinatorial optimization problem consists in incorporating to the problem an objective function, the constraints of the problem and assigning values to variables, which must be defined to determine the set of solutions respecting the constraints.

The function M(x), as given in formula (1), is composed of three variables namely: time, cost and load. It comes to a problem that integrates multiple criteria. These criteria, being often contradictory, can render resolving the problem more difficult.

Then, we have recourse to a multi-objective optimization process, which consists of simultaneously optimize multiple objective functions. Our work proposes to adapt the branch and bound algorithm for solving this sort of problem. Within the implementation of this algorithm we propose to decompose this function M(x) into three sub-objective functions: **fobj$_{Time}$**, **fobj$_{Cost}$** and **fobj$_{Load}$**.

We assume execute n tasks in m virtual machines. The algebraic formulation of the problem is as follows:

- P$_{ij}$ execution time of task j in the virtual machine i ;
- C$_{ij}$ cost of using resources;
- L$_{ij}$ load in the resource i running task j

Thus, a permutation matrix "X" is defined to ensure the assignment of a task to a single processor, such as:

$$\begin{cases} x_{ij} = 1, & \text{if task } j \text{ is executed on} \\ & \text{virtual machine } i \quad (9) \\ x_{ij} = 0, & \text{otherwise} \end{cases}$$

We seek to:

- minimize the execution time, which is defined as the greatest completion time of all tasks in all machines. From the formula (6) described in [8], we have :

$$fobj_{Time} = max_{1 \leq i \leq m} \sum_{j=1}^{n} P_{ij} * x_{ij} \quad (10)$$

**VARIABLES :** V$_{Time}$= $\{r\_cpu, r\_comm\}$

**DOMAINS:**
$r\_cpu$ : $[20000, 50000]$ MIPS
$Lenght$ : $[100, 10000]$ MI
$File\ Size$ : $[35, 300]$ Ko
$Output\ Size$ : $[35, 300]$ Ko
$Bw$ : $[0, 10]$ MB/S

**CONSTRAINT:**
We can assume a maximum time not to be exceeded

- minimize the cost of using resources, which is defined as the summation cost of all tasks in all machines. the formulas are as following :

$$fobj_{Cost} \sum_{1 \leq i \leq m} \sum_{1 \leq j \leq n} C(i,j) * x_{ij} \quad (11)$$

**VARIABLES :**
V$_{Cost}$= $\{r_{cpu_{cost}}, r_{mem_{cost}}, r_{comm_{cost}}, rq_{cpu}, rq_{mem}, rq\_comm\}$

**CONSTRAINTS:**
Cost <= Budget set by the client

We can assume other constraints.

- minimize system load, which is defined as the largest load in all machines.

$$fobj_{Load} = max_{1 \leq i \leq m} \sum_{j=1}^{n} L_{ij} * x_{ij} \quad (12)$$

With:

$Load = 1 - [(1 - Load_{cpu})^{\omega L1}(1 - Load_{mem})^{\omega L2}(1 - Load_{bw})^{\omega L3}] \quad (13)$

**VARIABLES :** V$_{Load}$= $\{Load\_cpu, Load\_mem, Load\_br\}$

*C. Branch And Bound algorithm*

```
Function AlgoBnB
  Begin
  vmList : List<Vm>
  taskList : List<Task>
  BestSol ← Null : Initialize a list that will contain the
                   optimal solution
  constPopulationInitiale(vmList, taskList) : create initial
  population
  vm : VM : Virtual Machine v

    ubTime  ← ∞   ⎫
    ubcost  ← ∞   ⎬  initialize Upperbound
    ubLoad  ← ∞   ⎭     to a higher value
    lbTime  ← null
    lbcost  ← null  ⎫ initialize lowerbound
    lbLoad  ← null  ⎭      to zéro

  Function calcrecursif(vmList : List<Vm>, currentbest :
  List<Vm>): this method selects (for each level) a node
           having the smallest evaluation as vertex to explore

  vmlistlocal ← vmList.clone()
  cureentbestlocal ← currentbest.clone()

  IF(vmListlocal.size()==1)
      calctrier(vmListlocal, v) : Calculate lower bound of
      the machine and check if the terminal is improved:

      IF(lbTime<ubTime && lbcost<ubCost &&
            lbLoad<ubLoad) then

          ubTime  ← lbTime    ⎫  Updating optimal
          ubcost  ← lbCost    ⎬  solution and its
          ubLoad  ← lbLoad    ⎪    terminals
          BestSol ← currentbestlocal  ⎭  upperbound
      EndIF
  EndIF
  Else
      calctrier(vmListlocal, vm)  ⎫  Calculate lower
                                   ⎬  bound of each vertex
                                   ⎪  and check if there
                                   ⎭  are improvement

  test if evaluation of the node is better than the current
  evaluation
```

```
      IF(lbTime<ubTime && lbcost<ubCost &&
         lbLoad<ubLoad) then
             currentbestlocal ← vm : Add the partial
             solution in currentbestlocal
             subVms ← genererSublist() : generate new
             subset which does not contain the processed
             node
             calcrecursif(subVms, currentbestlocal) :
      EndIF
      Else
          prune the node; return
  End
EndFunction
```

VI. EXPERIMENTATION AND RESULTS

We implemented in Java our approach based on the branch-and-bound applied to the task scheduling, in cloud computing and considering QoS constraints. We compared the experimentation results with those of the genetic algorithm. To do so, we randomly generated a number of 4 virtual machines and a number of tasks from a range of values {10, 100}. Table 1 and Table 2 show the configuration of the resources. We put the weighting coefficients with values between 0 and 1, corresponding to the user's cpu rate, memory, and bandwidth according to our needs.

For the calculation of resource utilization costs, we proposed a billing formula per unit. For each second of CPU time, and for each 1024 MB occupied, and for every 100 GB of data space, and for each 1MB/s taken of the bandwidth; will be billed just one unit cost.

TABLE I. HOST CONFIGURATION

| N° of Host | 1 |
|---|---|
| Processing Power (MIPS) | 150 000 |
| RAM (MB) | 256 000 |
| Bw (Mb/s) | 2000 |

TABLE II. VIRTUAL MACHINES CONFIGURATION

| Virtual machines | VM1 | VM2 | VM3 | VM4 |
|---|---|---|---|---|
| P.Power (MIPS) | 1024 | 4096 | 4096 | 4096 |
| RAM (MB) | 4000 | 3000 | 5000 | 5000 |

The results of this experimentation are shown in Figure 1, Figure 2 and Figure 3.

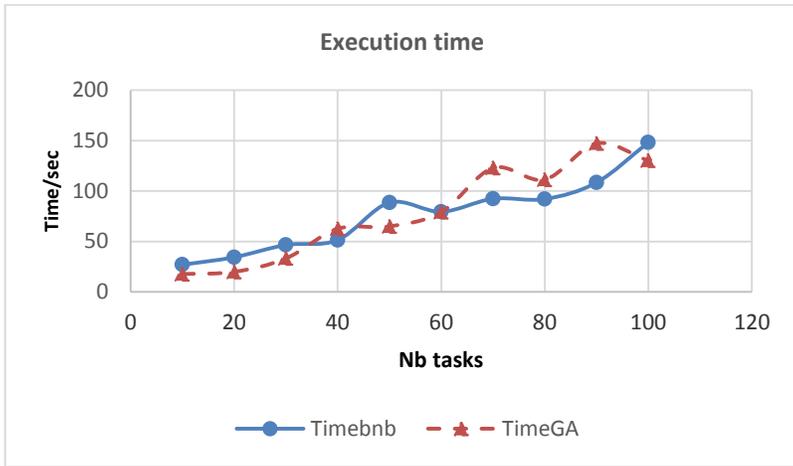

Fig. 1. Execution time graphics

Figure 1 illustrating the evolution of time according to the number of tasks, shows that branch and bound algorithm (BnB) gives better results. The graphic increases gradually. For a number of tasks lower than 30, the genetic algorithm is slightly better. The intersection of the two graphics appears from this point, where time BnB has a tendency to decrease. Our model gives a good prediction; and becomes better beyond 30 tasks than the genetic algorithm.

The Figure 2 for cost shows that the curve is nearly similar to the Figure 1. The curves of time intersect at 35 tasks. From this point, the cost of BnB is less than that of the genetic algorithm. This proves that BnB can ensure optimal solutions and better satisfy cloud computing users.

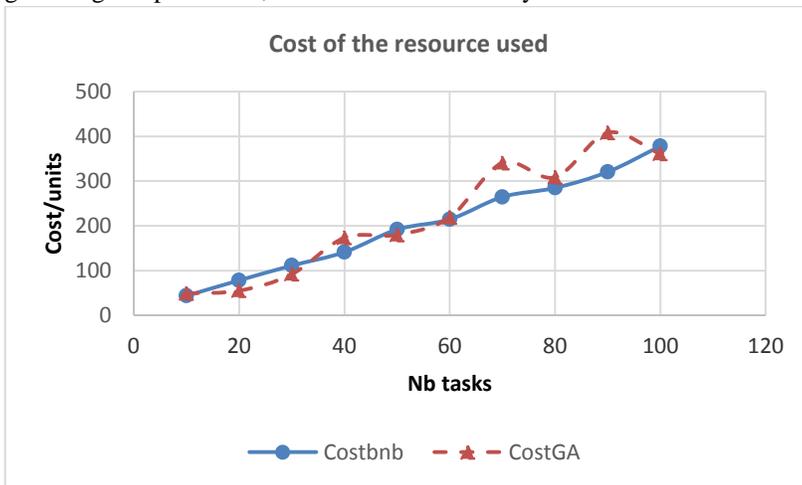

Fig. 2. Cost graphics

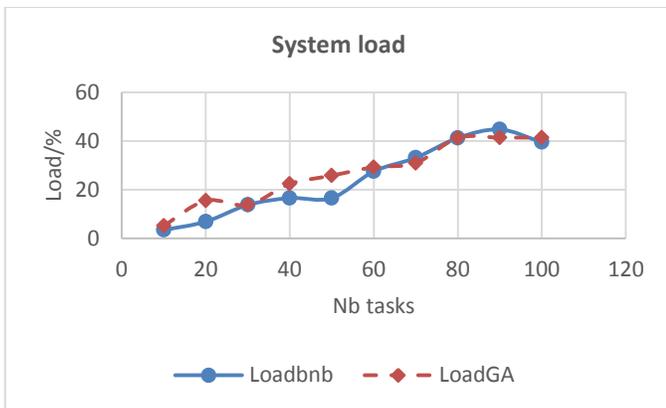

Fig. 3. Load graphics

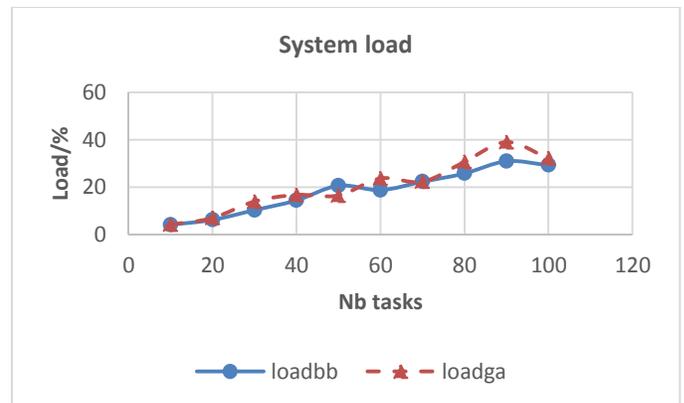

Fig. 4. Load graphics

The system load is a constraint in the cloud. It can unbalance the resource scheduling. We can deduce from Figure 3, that the system load with BnB is improved. The load increases more and more when increasing the number of tasks, the graphics show that the BnB consumes less load compared to the genetic algorithm. Other tests were performed. Among the obtained results, the graph of Figure 4, confirms that the BnB algorithm remains better and consumes less load though we increase the number of tasks.

Therefore, we can conclude that both genetic and BnB algorithm can solve optimization problems effectively and better meet the requirements of quality of service. Thus, when it is a small population of tasks, we can opt for the BnB algorithm, and for a population from a hundred tasks, the genetic model is more suitable and outperforms the BnB algorithm approach. Finally, the virtual machine is selected based on the cost and processing power.

From these comparisons, we can conclude that the Branch and Bound algorithm can solve optimization problems effectively and better meet the requirements of quality of service. Thus, when there is a small population of nodes, we can choose the BnB algorithm, and for a large population, the genetic model is more suitable and surpasses the BnB approach. Finally, all these results allow us to determine the capacity of the branch and bound algorithm, which is limited to a number of nodes not exceeding 12. Beyond that, the running time increases rapidly and can cause program shutdown.

## VII. CONCLUSION

In this paper, we studied an algorithm for task scheduling in a cloud computing environment using an exact approach. Our approach is based on the branch-and-bound algorithm, incorporating the QoS constraints on both aspects: the user aspect and the aspect of system load balance. We followed a rational approach, based on a comparative study with the focus on the value analysis. We showed the interest of this algorithm through experimental results that allowed us to evaluate the performance of cost and system load. In comparison with the genetic algorithm, the Branch-and-Bound technique gave better results for a small population of nodes in terms of time, cost and load balancing. We can conclude that the establishment of an effective task scheduling system can meet the requirements of users, with a good use of resources, and improving the overall performance of the cloud computing environment. Thus, scheduled tasks are regarded as a management tool for cloud servers.